\documentclass[onecolumn,preprintnumbers,eqsecnum,superscriptaddress,11pt]{revtex4}
\usepackage{amsfonts}
\usepackage{amssymb}
\usepackage{amsmath}
\usepackage{epsf}
\usepackage{graphicx}
\usepackage{dcolumn}
\usepackage{bm}
\usepackage{CJK}
\usepackage{color}
\usepackage[colorlinks=true,
            citecolor=blue,
            urlcolor=blue,
            filecolor=black,
            linktocpage=true,
            linkcolor=blue,]{hyperref}

\def\nn{{\nonumber}}

\newcommand{\beq}{\begin{equation}}
\newcommand{\eeq}{\end{equation}}
\newcommand{\beqs}{\begin{eqnarray}}
\newcommand{\eeqs}{\end{eqnarray}}

\newcommand{\be}{\begin{equation}}
\newcommand{\ee}{\end{equation}}
\newcommand{\bea}{\begin{eqnarray}}
\newcommand{\eea}{\end{eqnarray}}

\def\tildeF{\tilde F}

\def\Chone{C_{h1}}
\def\Chtwo{C_{h2}}
\def\Ckone{C_{k1}}
\def\Cktwo{C_{k2}}
\def\Cavone{C_{a1}}
\def\Cavtwo{C_{a2}}
\def\Cthree{C_{j1}}
\def\Cfour{C_{j2}}
\def\CTHREE{C_{j1}}
\def\CFOUR{C_{j2}}

\def\[{\left[}
\def\]{\right]}
\def\({\left(}
\def\){\right)}

\begin{document}

\title{Bulk Viscosity of dual Fluid at Finite Cutoff Surface via Gravity/Fluid correspondence in Einstein-Maxwell Gravity}

\author{Ya-Peng Hu}\email{huyp@nuaa.edu.cn}
\address{College of Science, Nanjing University of Aeronautics and Astronautics, Nanjing 211106, China}
\address{INPAC, Department of Physics, and Shanghai Key Lab for Particle Physics and Cosmology, Shanghai Jiao Tong University, Shanghai 200240, China}
\address{State Key Laboratory of Theoretical Physics, Institute of Theoretical Physics, Chinese Academy of Sciences, Beijing 100190, China}

\author{Yu Tian}\email{ytian@ucas.ac.cn}
\address{School of Physics, University of Chinese Academy of Sciences, Beijing 100190, China}
\address{State Key Laboratory of Theoretical Physics, Institute of Theoretical Physics, Chinese Academy of Sciences, Beijing 100190, China}

\author{Xiao-Ning Wu}\email{wuxn@amss.ac.cn}
\address{Institute of Mathematics, Academy of Mathematics and System Science, Chinese Academy of Sciences, Beijing 100190, China}
\address{State Key Laboratory of Theoretical Physics, Institute of Theoretical Physics, Chinese Academy of Sciences, Beijing 100190, China}
\address{Hua Loo-Keng Key Laboratory of Mathematics, CAS, Beijing 100190, China}

\begin{abstract}
Based on the previous paper arXiv:1207.5309, we investigate the possibility to find out the bulk viscosity of dual fluid at the
finite cutoff surface via gravity/fluid correspondence in
Einstein-Maxwell gravity. We find that if we adopt new conditions to
fix the undetermined parameters contained in the stress tensor and
charged current of the dual fluid,
two new terms appear in the stress tensor of the dual fluid. One new term is related to the bulk
viscosity term, while the other can be related to the perturbation of
energy density. In addition, since the parameters contained in the
charged current are the same, the charged current is not changed.


\end{abstract}

\maketitle

\vspace*{1.cm}

\newpage

\section{Introduction}

The AdS/CFT
correspondence~\cite{Maldacena:1997re,Gubser:1998bc,Witten:1998qj,Aharony:1999ti}
provides a remarkable connection between a gravitational theory and
a quantum field theory. According to the correspondence, the
gravitational theory in an asymptotically AdS spacetime can be
formulated in terms of a quantum field theory on its boundary. In
particular, the dynamics of a classical gravitational theory in the
bulk is mapped to a strongly coupled quantum field theory on the
boundary. Therefore, AdS/CFT provides a useful tool and some
insights to investigate the strongly coupled field theory from the
dual classical gravitational
theory~\cite{Herzog:2009xv,CasalderreySolana:2011us}.

Since the discovery of the AdS/CFT correspondence, there has been
much work studying the hydrodynamical behavior of the dual quantum
field theory using this
correspondence~\cite{Policastro:2001yc,Kovtun:2003wp,Buchel:2003tz,Kovtun:2004de}, and a simple reason is that the hydrodynamics can be an effective description of any
interacting quantum field theory in the long wave-length limit, i.e.
when the length scales under consideration are much larger than the
correlation length of the quantum field theory. Furthermore, as the long wave-length limit
of the AdS/CFT correspondence, the gravity/fluid correspondence has also been proposed in~\cite{Bhattacharyya:2008jc}. The
big advantage of the gravity/fluid correspondence is that it can provide a systematic
way to map the boundary fluid to the bulk gravity. Besides the first
order stress-energy tensor, the charged current of dual fluid can
be both obtained by using the gravity/fluid
correspondence~\cite{Hur:2008tq, Erdmenger:2008rm, Banerjee:2008th,
Son:2009tf, Tan:2009yg, Torabian:2009qk, Hu:2010sn}. Furthermore, it
has been shown that the chiral magnetic effect (CME) and  the chiral
vortical effect (CVE) can be brought into the hydrodynamics via this
correspondence after adding the Chern-Simons term of Maxwell field
in the action~\cite{Son:2009tf, Tan:2009yg, Hu:2011ze,
Kalaydzhyan:2011vx, Amado:2011zx}.

Note that, the dual field theory in AdS/CFT correspondence or
gravity/fluid correspondence usually resides on the boundary with
infinite radial coordinate, and the corresponding dual field or
fluid are conformal. In fact, the AdS/CFT correspondence can also be
used to studying non-conformal fluids. A simple way of achieving
this is to break the conformal symmetry by introducing a finite
cutoff on the radial coordinate in the bulk, and it has been shown
that a Navier-Stokes (NS) fluid can live on the cutoff
surface $r=r_c$, which implies a deep relationship between the NS
equations and gravitational
equations~\cite{Bredberg:2010ky,Bredberg:2011jq,Compere:2011dx,Cai:2011xv,Niu:2011gu,Kuperstein:2011fn,Compere:2012mt,Kuperstein:2013hqa}.
Moreover, from the renormalization group (RG) viewpoint, the radial
direction of the bulk spacetime corresponds to the energy scale of
the dual field
theory~\cite{Balasubramanian:1998de,Akhmedov:1998vf,de
Boer:1999xf,Susskind:1998dq}. The infinite boundary corresponds to
the UV fixed point of the dual field theory, and hence cannot be
reached by experiments. Therefore, the physics at a finite cutoff
surface $r=r_c$ which means a finite energy scale becomes important,
and the dependence of transport coefficients of dual fluid on the
cutoff surface $r_c$ can be interpreted as the RG flow. In the
literature, there exist several approaches investigating the RG
flow, such as the Wilsonian fluid/gravity~\cite{Bredberg:2010ky},
the holographic Wilsonian RG~\cite{Heemskerk:2010hk,Faulkner:2010jy}
and the sliding membrane~\cite{Iqbal:2008by}. It has been found that
these apparently different approaches can be
equivalent~\cite{Sin:2011yh}. In addition, following the spirit of
the gravity/fluid correspondence
\cite{Bredberg:2010ky,Cai:2011xv,Kuperstein:2011fn}, the
investigation of the dual fluid on infinite boundary has also been
generalized to a finite cutoff surface, which contains the
Chern-Simons term of the Maxwell field in the
bulk~\cite{Bai:2012ci}. A little difference from the case with
infinite boundary, the first order stress-energy tensor and charged current of
the dual fluid at the finite cutoff surface contain
several undetermined parameters which relate to the boundary
conditions and gauge conditions. In order to fix these parameters,
the Dirichlet boundary condition and Landau frame have been chosen
in Ref~\cite{Bai:2012ci}. It has been found that
the dual fluid on the hypersurface is non-conformal, which is expected from the fact that the conformal symmetry has been broken with a finite
radial coordinate in the bulk. The shear
viscosity takes the same value as that at the infinite boundary~\cite{Banerjee:2012iz,Jensen:2012jy,Bardeen:1984pm}.
However, the dual stress tensor has zero bulk viscosity.
Usually, nonzero bulk viscosity can also break the conformal symmetry of dual
fluid. Therefore, how the nonzero bulk viscosity of dual fluid can appear at the finite
cutoff surface is the main issue focused on in this letter. A simple result is that the bulk viscosity can be obtained if we
let the energy density of fluid perturb, which means relaxing the usually chosen Landau
frame condition, i.e. $T^{(1)}_{v v}\neq0$.

The rest of the paper is organized as follows. In Sec.~II, we
present a review containing the main results in~\cite{Bai:2012ci}.
Particularly, we point out the Dirichlet boundary condition and
Landau frame to fix the undetermined parameters, and obtain the
dependence of transport coefficients on the radial cutoff $r_c$. In
Sec.~III, a new conditions are chosen to fix the undetermined
parameters, and the bulk viscosity can appear under this new conditions. Sec.~VII is devoted to the conclusion and discussion.

\section{Review: Holographic charged fluid at finite cutoff surface}
In this section, we will give a simple review to show the main
results in~\cite{Bai:2012ci}, which uses the gravity/fluid
correspondence to generalize the dual charged fluid on the infinite (conformal)
boundary to the finite cutoff surface, with a five-dimensional
Einstein-Maxwell gravity with Chern-Simons term in the bulk.

The action of the five-dimensional Einstein-Maxwell gravity with
Chern-Simons term is
\begin{equation}
\label{IVaction1} I=\frac{1}{16 \pi G}\int_\mathcal{M}~d^5x
\sqrt{-g^{(5)}} \left(R-2 \Lambda
\right)-\frac{1}{4g^2}\int_\mathcal{M}~d^5x
\sqrt{-g^{(5)}}(F^2+\frac{4\kappa_{cs} }{3}\epsilon
^{LABCD}A_{L}F_{AB}F_{CD})\ .
\end{equation}
The equations of motion are
\begin{eqnarray}
\label{IVeqs1}
R_{AB } -\frac{1}{2}Rg_{AB}+\Lambda g_{AB}-\frac{1}{2g^2}\left(F_{A C}{F_{B }}^{C}-\frac{1}{4}g_{AB}F^2\right)&=&0~, \\
\nabla_{B} {F^{B}}_{A}-\kappa_{cs} \epsilon_{ABCDE}F^{BC}F^{DE}
&=&0\ .\nonumber~~
\end{eqnarray}
and the starting point is the five-dimensional charged RN-AdS black
brane solution \cite{Cai:1998vy,Cvetic:2001bk,Anninos:2008sj}
\begin{eqnarray}
ds^2=\frac{dr^2}{r^2f(r)}+r^2
 \left(\mathop\sum_{i=1}^{3}dx_i^2 \right)-r^2f(r) dt^2, \label{IVSolution}
\end{eqnarray}%
where
\begin{eqnarray}
\label{IVf-BH} f(r) &=& 1-\frac{2M}{r^{4}}+\frac{Q^2}{r^6},~~F =
-g\frac{2\sqrt 3 Q}{r^3}dt \wedge dr\ . ~~
\end{eqnarray}%
Note that, the RN-AdS black brane solution still solves the equation
(\ref{IVeqs1}) even in the presence of Chern-Simons term.
From~(\ref{IVSolution}), the outer horizon of the black brane is
located at $r=r_{+}$, where $r_{+}$ is the largest root of $f(r)=0$,
and its Hawking temperature is
\begin{eqnarray}
T_{+}&=&\frac{(r^2f(r))'}{4 \pi}|_{r=r_{+}}=\frac{1}{2 \pi
r_{+}^3}(4M-\frac{3Q^2}{r_{+}^2})\ .\label{IVTemperature}
\end{eqnarray}
In order to avoid the coordinate singularity, one can write the
above black brane solution in the Eddington-Finkelstin coordinate
system
\begin{eqnarray}\label{IVSolution1}
ds^2 &=& - r^2 f(r)dv^2 + 2 dv dr + r^2(dx^2 +dy^2 +dz^2)\ ,\\
F &=& -g\frac{2\sqrt 3 Q}{r^3}dv \wedge dr\ , \notag~~
\end{eqnarray}
where $v=t+r_*$, and $r_*$ is the tortoise coordinate satisfying
$dr_*=dr/(r^2f)$.

Since the holographic charged fluid is considered at some cutoff
hypersurface $r=r_c$ ($r_c$ is a constant), thus it is useful to
make a coordinate transformation $v\rightarrow v/\sqrt{f(r_c)}$ in
the solution (\ref{IVSolution1}), and the simple reason is explicitly making the
induced metric on the cutoff surface conformal to flat metric, i.e. the cutoff surface with metric
$ds^2=r_{c}^2 (-dv^2+dx^2+dy^2 +dz^2)$. It should be pointed that
the Hawking temperature is expressed as $T=T_{+}/\sqrt{f(r_c)}$ with respect to the killing observer $(\partial/\partial v)^a$ in the new
coordinate system, and the RN-AdS black brane solution becomes
\begin{eqnarray}\label{IVSolution2}
ds^2 &=& - \frac{r^2 f(r)}{f(r_c)}dv^2 + \frac{2}{\sqrt{f(r_c)}} dv dr + r^2(dx^2 +dy^2 +dz^2)\ ,\\
F &=& -g\frac{2\sqrt 3 Q}{r^3 \sqrt{f(r_c)}}dv \wedge dr\ . \notag~~
\end{eqnarray}
Since we expect to obtain the transport coefficients of dual fluid at finite cutoff surface like shear viscosity $\eta$, which usually needs the non-constant velocity of fluid. Therefore, we can first give the above static black brane a constant velocity through a boosted transformation, which obtains the five-dimensional boosted RN-AdS black brane solution
\begin{eqnarray}   \label{IVrnboost}
ds^2 &=& - \frac{r^2 f(r)}{f(r_c)}( u_\mu dx^\mu )^2 - \frac{2}{\sqrt{f(r_c)}} u_\mu dx^\mu dr + r^2 P_{\mu \nu} dx^\mu dx^\nu, \\
A&=&\frac{\sqrt 3 g Q}{r^2
\sqrt{f(r_c)}}u_{\mu}dx^{\mu}, \notag \label{IVExternalfield}~~
\end{eqnarray}
with
\begin{equation}
u^v = \frac{1}{ \sqrt{1 - \beta_i^2} },~~u^i = \frac{\beta_i}{
\sqrt{1 - \beta_i^2} },~~P_{\mu \nu}= \eta_{\mu\nu} + u_\mu u_\nu\ .
\label{IVvelocity}
\end{equation}
where $x^\mu=(v,x_{i})$ is the boundary coordinates, velocities $\beta^i $ are constants,
$P_{\mu \nu}$ is the projector onto spatial directions, and the boundary indices $(\mu,\nu)$ are raised and lowered with the Minkowsik
metric $\eta_{\mu\nu}$, while the bulk indices have been distinguished by $(A,B)$. Note that, after boosting the black brane solution in Eq.~(\ref{IVSolution2}), the metric~(\ref{IVrnboost}) describes a
uniform boosted black brane moving at velocity $\beta^i $, and one
obtains a solution with more parameters which can then be related to
the degrees of freedom of the dual boundary
fluid~\cite{Bhattacharyya:2008jc}. In addition, another underlying idea is that the boosted black brane solution will be related to a dual fluid solution of constant velocity.

As we know, if one expects to obtain the transport coefficients like shear viscosity $\eta$ of fluid, the fluid should be with non-constant velocity. And a simple way of constructing the fluid with non-constant velocity is to perturb the fluid away from equilibrium, i.e. promoting the constant velocity to $x_i$ and $v$-dependent functions, then one can find the equations of motion satisfied by fluid perturbation. On the dual gravity side,
this can be achieved by promoting the parameters in the boosted
black brane solution (\ref{IVrnboost}) to functions of boundary
coordinates $x^\mu$~\cite{Bhattacharyya:2008jc,Hur:2008tq}. Since the parameters now depend on the boundary
coordinates, the solution~(\ref{IVrnboost}) is no longer a solution of
the equation of motion~(\ref{IVeqs1}), extra correction terms are
needed to make (\ref{IVrnboost}) with non-constant parameters to be a solution. It is useful and convenient to define the
following tensors
\begin{eqnarray}
&&W_{AB} = R_{AB} + 4g_{AB}+\frac{1}{2g^2}\left(F_{AC}{B^{C}}_B +\frac{1}{6}g_{AB}F^2\right),\label{IVTensors1}\\
&&W_A = \nabla_{B} {F^{B}}_{A}-\kappa_{cs}
\epsilon_{ABCDE}F^{BC}F^{DE}~~,\label{IVTensors2}
\end{eqnarray}
where the convention $\epsilon_{vxyzr}=+\sqrt{-g}$ has been used. Note that, the right hand side of Eq.(\ref{IVTensors1}) is in fact equivalent to the left hand side of Eq.~(\ref{IVeqs1}), i.e. instead of the Ricci scalar with the Maxwell filed in Eq.~(\ref{IVeqs1}). Therefore, the solutions of equation motions should satisfy $W_{AB}=0$ and $W_{A}=0$. In addition, when the parameters
become functions of boundary coordinates $x^\mu$, $W_{AB}$ and $W_{A}$ no
longer vanish for (\ref{IVrnboost}) with non-constant parameters, and are proportional to the derivatives of the
parameters. These terms are the source terms which will be canceled
by extra correction terms introduced into the metric and Maxwell field. It should be pointed out that there is a trick to obtain the extra correction terms. According to the Refs \cite{Bhattacharyya:2008jc,Hur:2008tq}, one can first obtain the extra correction terms at the origin $x^{\mu}=0$. Then, the extra correction terms at any point can be simply obtained by making the extra correction terms at the origin $x^{\mu}=0$ into a covariant form. More details, in order to obtain the extra correction terms at the origin $x^{\mu}=0$, one can first expand
the parameters around $x^\mu=0$ to the first order
\begin{eqnarray}
\beta_i(x^\mu)&=&\partial_{\mu} \beta_{i}|_{x^\mu=0}
x^{\mu},~M(x^\mu)=M(0)+\partial_{\mu}
 M|_{x^\mu=0} x^{\mu},~Q(x^\mu)=Q(0)+\partial_{\mu}
Q|_{x^\mu=0} x^{\mu}. \label{IVExpand}
\end{eqnarray}
where $\beta^i(0)=0$ have been assumed at the origin, and we only consider the first order case in our manuscript. Obviously, by construction here $x^\mu$ are small quantities which can be counted like $\epsilon$. Moreover, since the power of $x^\mu$ and the order of derivatives are always equal in every term of the Taylor expansion in (\ref{IVExpand}), the small $x^\mu$ expansion is equivalent to the relativistic hydrodynamic limit which is frequently taken in the fluid/gravity literatures, i.e. $\partial_v \sim \epsilon$ and $\partial_i \sim \epsilon$. Therefore, the first
order source terms can be obtained by inserting the solution (\ref{IVrnboost})
with~(\ref{IVExpand}) into $W_{AB }$ and $W_{A}$.
By choosing an appropriate gauge like the background field gauge in \cite{Bhattacharyya:2008jc} ($G$
represents the full metric)
\begin{equation}
G_{rr}=0,~~G_{r\mu}\propto
u_{\mu},~~Tr((G^{(0)})^{-1}G^{(1)})=0\ ,\label{gauge}
\end{equation}
and taking into account the spatial $SO(3)$
symmetry preserved in the background metric~(\ref{IVSolution2}), the
first order correction terms around $x^\mu=0$ needed to compensate
for the source terms are \cite{Bhattacharyya:2008jc}
\begin{eqnarray}\label{IVcorrection}
&&ds_{(1)}^2 = \frac{ k(r)}{r^2}dv^2 +
2\frac{h(r)}{\sqrt{f(r_c)}}dv dr + 2 \frac{j_i(r)}{r^2}dv dx^i
 +r^2 \left(\alpha_{ij}(r) -\frac{2}{3} h(r)\delta_{ij}\right)dx^i dx^j, \\
&&A^{(1)} = a_v (r) dv + a_i (r)dx^i~~.\label{IVcorrection1}
\end{eqnarray}

Note that, the region interested in is between the outer
horizon and cutoff surface $r_{+}\leq r \leq r_c$. By requiring a
cancellation between the effects of correction and source terms, one
obtains
\begin{eqnarray}
h(r)&=&{\Chtwo}+\frac{{\Chone}}{r^4},\nn\\
a_v(r)&=&{\Cavtwo}+\frac{\Cavone}{r^2}-\frac{2 {\Chone} g Q}{\sqrt{3} r^6 \sqrt{f\left(r_c\right)}},\nn\\
k(r)&=&{\Cktwo}+{\Ckone} r^2-\frac{2 {\Chtwo} r^4}{f\left(r_c\right)}+\frac{4 {\Chone} \left(-Q^2+M r^2\right)}{3 r^6 f\left(r_c\right)}+\frac{2 \Cavone Q}{\sqrt{3} g r^2 \sqrt{f\left(r_c\right)}}+\frac{2 r^3 \partial _i\beta _i}{3 \sqrt{f\left(r_c\right)}},\nn\\
\alpha_{ij}(r)&=&\alpha(r)\left\{(\partial_i \beta_j + \partial_j
\beta_i )-\frac{2}{3} \delta_{ij}\partial_k \beta^k \right\}\ ,
\end{eqnarray}
where $\alpha(r)$ is
\begin{eqnarray}
\alpha(r)&&= \int_{r_c }^{r}\frac{s^{3}-r_{+}^3}{-s^{5}f(s)}\sqrt{f(r_c)}ds\ ,
\end{eqnarray}
and the Dirichlet boundary condition at $r=r_c$, regularity of the
bulk fields at the future horizon have been used to obtain
$\alpha(r)$. In addition, $h(r)$ is solved from $W_{rr}=0$, while
$a_v(r)$ is solved from $W_{r}=0$ or $W_{v}=0$, and $k(r)$ is solved
from $W_{vv}=0$ or $W_{vr}=0$. $j_i(r)$ and $a_i(r)$ are more
difficult to solve since they couple with each other. For details on
how to solve their equations see~\cite{Bai:2012ci}. Therefore, the extra correction terms at the origin are obtained, and hence the extra correction terms at any point of the whole cutoff surface can be constructed from the extra correction terms at the origin by making them into a covariant form~\cite{Bhattacharyya:2008jc,Hur:2008tq}, which finally gives the first order perturbative solution for the bulk.

Given the first order perturbative solution for the bulk, we are able to extract information of the dual fluid using the gravity/fluid correspondence. According to the correspondence, the stress tensor $T_{\mu\nu}$ of dual fluid residing at the cutoff surface with the induced metric $\gamma_{\mu\nu}$ is given by~\cite{Bredberg:2011jq,Balasubramanian:1999re,de
Haro:2000xn,Emparan:1999pm,Mann:1999pc}
\begin{equation}
 T_{\mu\nu}=2\left(K_{\mu\nu}-K\gamma _{\mu\nu}-C\gamma _{\mu\nu}\right)\ , \label{IVTabCFT}
\end{equation}
where $\gamma_{\mu\nu}$ is the boundary metric obtained from the well-known ADM decomposition
\begin{eqnarray}
ds^2 = \gamma_{\mu\nu}(dx^\mu + V^\mu dr)(dx^\nu + V^\nu dr) + N^2
dr^2\ ,
\end{eqnarray}
the extrinsic curvature is $K_{\mu\nu}=-\frac{1}{2}(\nabla_{\mu}n_{\nu}+\nabla_{\nu}n_{\mu})$, and $n^{\mu}$ is the unit normal vector of the constant hypersurface
$r=r_c$ pointing toward the $r$ increasing direction. Note that, here and hereafter we mainly investigate the stress tensor of dual fluid $T_{\mu\nu}$ residing on the spacetime $\gamma _{\mu\nu}$, which will be found to be more consistent with the next section and sufficient for the discussion. In addition, the term $C\gamma _{\mu\nu}$ is usually related to the boundary counterterm added to cancel the divergence of the stress tensor $T_{\mu\nu}$ when the boundary $r=r_c$ approaches to infinity, for
example $C=3$ in the asymptotical $AdS_5$ case. However, in our case
with finite boundary there is no divergence of the stress
tensor, the reason that we still add the boundary counterterm is to
compare with the results obtained for $r_c$ approaches infinity.

For the dual charged current at the finite cutoff surface, it can be
computed via
\begin{equation}\label{IVcurrent}
J^\mu = \lim_{r \rightarrow r_c}r_{c}^4\frac{1}{\sqrt{-\gamma}}
\frac{\delta S_{cl}}{\delta \hat A_\mu} = - \lim_{r \rightarrow r_c}
r_{c}^4 \frac{N}{g^2} (F^{r \mu}+\frac{4\kappa_{cs} }{3}\epsilon
^{r\mu \rho \sigma \tau }A_{\rho}F_{\sigma \tau })~,
\end{equation}
where $\hat A_\mu$ is the gauge field $A_\mu$ projected to boundary.

It is obvious that there are nine parameters $\Chone,\Chtwo, \Ckone,
\Cktwo, \Cavone, \Cavtwo, \Cthree, \Cfour$ and $C$ in the stress
tensor and charged current of the dual fluid, where the solution of
$j_i(r)$ contains two constants $\CTHREE$ and $\CFOUR$. In order to
extract useful information on the dual fluid, one needs to fix these
parameters. In Ref~\cite{Bai:2012ci}, after considering the
consistency which deduces $\Ckone=0$, $C=3$ for the comparison. The Dirichlet boundary condition has also been chosen like\beqs
    h(r_c) = 0 ,\quad k(r_c) = 0,\quad j_i(r_c) = 0,\quad a_v(r_c) = 0,\quad a_i(r_c) =
    0,~\label{DirichletB}
\eeqs and the underlying physics of this boundary condition is to keep the induced metric $\gamma_{\mu\nu}$ to be flat metric or conformal to flat metric, thus has a well-defined boosted transformation at the cutoff surface. In addition, the following conditions are also chosen
\begin{equation}
T^{(1)}_{v v} =0, ~~T^{(1)}_{v x} =0,~~J_{(1)}^{v}
=0.~\label{LandauFrame}
\end{equation}
where $T^{(1)}_{v x} =0$ is a gauge choice which relates to the fact that one can always choose a suitable coordinate system to make the fluid with zero velocity. In addition, the physics of other two conditions can relate to assume that there are no perturbation of energy density and charge density at the cutoff surface. Therefore, excluding $\Ckone$ and $C$, there are seven independent parameters that can be set freely. Note that, $j_i(r_c) = 0$ and $a_i(r_c)=0$ are in fact one equation since $a_i(r)$ can be obtained from $j_i(r)$ as shown in \cite{Bai:2012ci}. Hence, these seven independent parameters can be just totally fixed by the boundary conditions in (\ref{DirichletB}) and (\ref{LandauFrame}).

Therefore, the nine parameters can be finally fixed as \beqs
    & &{\Chone}=-\frac{\partial _i\beta _i r_c^3}{4 \sqrt{f\left(r_c\right)}},\quad{\Chtwo}=\frac{\partial _i\beta _i}{4 \sqrt{f\left(r_c\right)} r_c},\quad{\Ckone}=0,\nn\\
    & &\Cavone=-\frac{\sqrt{3} g Q \partial _i\beta _i}{2 f\left(r_c\right) r_c},\quad{\Cavtwo}=\frac{g Q \partial _i\beta _i}{\sqrt{3} f\left(r_c\right) r_c^3},\quad{\Cktwo}=-\frac{\partial _i\beta _i \left(-10 M+r_c^4\right)}{6 f\left(r_c\right){}^{3/2} r_c},\nn\\
    & &{\Cthree}=0,\quad{\Cfour}=\partial _v\beta _x r_c^3 \left(-1+\sqrt{f\left(r_c\right)}\right)+\frac{Q \partial _xQ-\partial _xM r_c^2}{r_c^3 \sqrt{f\left(r_c\right)}},~{C}=3\ .
\eeqs

Consequently, the non-zero components of $T_{\mu\nu}$, i.e. the energy
stress tensor of dual fluid residing at the cutoff surface with the induced metric $\gamma_{\mu\nu}$, are \beqs
    & &T^{(0)}_{v v} = 6\left(1- \sqrt{f\left(r_c\right)} \right) r_c^2, \quad T^{(0)}_{i i} =\frac{-4 M+6\left(1- \sqrt{f\left(r_c\right)}\right) r_c^4}{r_c^2\sqrt{f\left(r_c\right)}},\nn\\
   & &T^{(1)}_{i j} = - 2r_+^3\sigma _{ij}/r_c^2\ ,
\eeqs
which can be further rewritten into a covariant form
\begin{eqnarray}
T_{\mu \nu}=\rho\,U_\mu U_\nu+p
\Pi_{\mu\nu}-2\eta\sigma_{\mu\nu},
\label{IVStressTensor1}
\end{eqnarray}
where
\begin{eqnarray}
&&U_{\mu}=r_c u_{\mu},~~
\Pi_{\mu\nu}=\gamma_{\mu\nu}+U_{\mu}U_{\nu}=r_c^2\eta_{\mu\nu}+U_{\mu}U_{\nu}, \nonumber\\
&&\sigma ^{\mu  \nu }\equiv \frac{1}{2} \Pi^{\mu  \alpha } \Pi^{\nu
\beta } \left(\nabla _{\alpha }U_{\beta } +\nabla _{\beta }U_{\alpha
}\right)-\frac{1}{3} \Pi^{\mu  \nu } \nabla _{\alpha }U^{\alpha }.
\end{eqnarray}
From (\ref{IVStressTensor1}), one can read out the energy density
$\rho$, pressure $p$ and shear viscosity $\eta$ of the dual fluid at
the cutoff surface
\begin{eqnarray}
\rho=6\left(1- \sqrt{f\left(r_c\right)} \right),~~~p=\frac{-4
M+6\left(1- \sqrt{f\left(r_c\right)}\right)
r_c^4}{r_c^4\sqrt{f\left(r_c\right)}}, ~~~\eta=r_+^3/r_c^3. \label{IVets}
\end{eqnarray}
Obviously, the dual fluid obtained at
the finite cutoff surface is indeed not conformal because the trace of $T_{\mu\nu}$ is nonzero, i.e. $\rho=3p$ has been broken. This result is consistent with that in Ref.~\cite{Kuperstein:2011fn}, and expected from the fact that the conformal symmetry has been broken with a finite
radial coordinate in the bulk. However, the bulk viscosity which can also break the conformal symmetry is absent.

The zeroth- and first-order charged current of the dual fluid are
given by \beqs
J_{(0)}^{\mu} &=& \frac{2\sqrt 3 Q}{g} u^\mu =: n u^\mu\,,\label{IVzeroth order current}\\
J_{(1)}^{\mu } &=&-\kappa P^{\mu \nu }\partial _{\nu}(\frac{\mu }{T}%
)+\sigma _{E}E^{\mu }+\sigma _{B}B^{\mu }+\xi \omega ^{\mu },
\label{IVfirst order current} \eeqs where $n$ is particle number
density and \beqs E^{\mu}&=& \tildeF^{\mu\nu }u_{\nu}, ~B^{\mu
}=\frac{1}{2}\epsilon ^{\mu \nu \rho \sigma}u_{\nu}\tildeF_{\rho
\sigma},~\omega ^{\mu }=\epsilon ^{\mu \nu \rho \sigma
}u_{\nu}\partial _{\rho}u_{\sigma}\,. \eeqs Note that, here
$\tildeF_{\mu \nu}$ is defined at the cutoff surface $r=r_c$ through
$A_{\mu}= \frac{\sqrt 3 g Q}{r_c^2
\sqrt{f(r_c)}}u_{\mu}$, and the chemical potential is defined as
\begin{eqnarray}
\mu =  A_v (r_c)-A_v (r_+)=\frac{\sqrt{3}g
Q}{\sqrt{f\left(r_c\right)}}\left(
\frac{1}{r_+^2}-\frac{1}{r_c{}^2}\right) ~,\label{IVchemical
potential}
\end{eqnarray}
where $M$, $Q$ and $r_+$ are not constants.

The transport coefficients are found to be \beqs \kappa &=&\frac{16
\pi ^2 r_+^7 T_+^3}{g^2 r_c^{10} \sqrt{f\left(r_c\right)}
f'(r_c)^2}\quad,~
 \sigma_E = \frac{16 \pi ^2 r_+^7 T_+^2}{g^2 r_c^{10} f'\left(r_c\right)^2},\nn\\
\sigma_B &=& -\frac{8 Q \left(3 r_c^2-2 r_+^2\right) \kappa _{\text{cs}}}{\sqrt{3} g r_+^2 r_c^2 \sqrt{f\left(r_c\right)}}+\frac{24 \sqrt{3} Q^3 \left(r_c^2-r_+^2\right){}^2 \kappa _{\text{cs}}}{g r_+^4 r_c^9 \sqrt{f\left(r_c\right)} f'\left(r_c\right)},\nn\\
\xi &=& -\frac{12 Q^2 \left(r_c^2-r_+^2\right){}^2 \kappa
_{\text{cs}}}{r_+^4 r_c^4 f\left(r_c\right)}+\frac{48 Q^4
\left(r_c^2-r_+^2\right){}^3 \kappa _{\text{cs}}}{r_+^6 r_c^{11}
f\left(r_c\right) f'\left(r_c\right)}\ , \eeqs which can reproduce the
results for infinite boundary if we take the limit $r_c$ approaches
infinity.

\section{The new conditions and bulk viscosity}
Note that, excluding $\Ckone$ and $C$, the stress tensor and charged current of dual fluid at
finite cutoff surface in fact depend on the seven parameters $\Chone, \Chtwo, \Cktwo, \Cavone, \Cavtwo, \Cthree$ and $\Cfour$, since
the stress tensor $T_{\mu\nu}$ is \beqs
    T^{(0)}_{v v} &=& 2\left(C-3 \sqrt{f\left(r_c\right)}\right) r_c^2,\nn\\
    T^{(0)}_{i i} &=&\frac{-4 M+2\left(3-C \sqrt{f\left(r_c\right)}\right) r_c^4}{\sqrt{f\left(r_c\right)} r_c^2},\nn\\
    T^{(1)}_{v v} &=&-2\partial _i\beta _i r_c+6 \sqrt{f\left(r_c\right)} r_c^2 h\left(r_c\right)+\frac{\left(-2 C+9 \sqrt{f\left(r_c\right)}\right) k\left(r_c\right)}{ r_c^2}+2\sqrt{f\left(r_c\right)} r_c^3 h'\left(r_c\right),\nn\\
    T^{(1)}_{v i} &=& -\frac{Q \partial _iQ}{ f\left(r_c\right) r_c^5}+\frac{\partial _iM}{f\left(r_c\right) r_c^3}-\partial _v\beta _i r_c\nn\\
    & &+2\frac{\left(-Q^2+\left(2-C \sqrt{f\left(r_c\right)}+3 f\left(r_c\right)\right) r_c^6\right) j_i\left(r_c\right)}{\sqrt{f\left(r_c\right)} r_c^8}-\frac{\sqrt{f\left(r_c\right)} j_i'\left(r_c\right)}{ r_c},\nn\\
    T^{(1)}_{i j} &=&2\left(2 \delta _{i j}\partial _k\beta _k-\partial _{(i}\beta _{j)} \right)r_c+2\delta _{i j}\frac{\partial _k\beta _k \left(2 M-3 r_c^4\right)}{3 f\left(r_c\right) r_c^3}\nn\\
    & & +2\left(-Cr_c^2+\frac{-2 M+3 r_c^4}{\sqrt{f\left(r_c\right)} r_c^2}\right)a_{i j}\left(r_c\right)-\sqrt{f\left(r_c\right)} r_c^3 a_{i j}'\left(r_c\right)\nn\\
    & & +2\delta _{i j}\left(\left(\frac{2 c r_c^2}{3}+\frac{5 \left(2 M-3 r_c^4\right)}{3 \sqrt{f\left(r_c\right)} r_c^2}\right) h\left(r_c\right)-\frac{2}{3} \sqrt{f\left(r_c\right)} r_c^3 h'\left(r_c\right)\right)\nn\\
    & & +2\delta _{i j}\left(\frac{\left(-2 M+\left(3-2 f\left(r_c\right)\right) r_c^4\right) k\left(r_c\right)}{2 \sqrt{f\left(r_c\right)} r_c^6}-\frac{\sqrt{f\left(r_c\right)} k'\left(r_c\right)}{2
    r_c}\right).\label{BoundaryST}
\eeqs while the charged current is \beqs
&&J^{v}_{(1)}=-\frac{2 \sqrt{3} Q h\left(r_c\right)}{g}+\frac{\sqrt{3} Q k\left(r_c\right)}{g r_c^4}+\frac{\sqrt{f\left(r_c\right)} r_c^3 a_v'\left(r_c\right)}{g^2},\nn\\
&&J_{(1)}^{i}=-\frac{2 \sqrt{3} Q j_i\left(r_+\right)}{g r_+^4}-\frac{\sqrt{3} Q \partial _v\beta _i}{g r_+ \sqrt{f\left(r_c\right)}}-\frac{\sqrt{3} Q \partial _iM}{g r_+ r_c^4 f\left(r_c\right){}^{3/2}}-\frac{\sqrt{3} \partial _iQ \left(-Q^2+r_c^6 f\left(r_c\right)\right)}{g r_+ r_c^6 f\left(r_c\right){}^{3/2}}\nn\\
&&~~~~~~~~~- \frac{4 \kappa _{\text{cs}} Q^2  \left(r_+^4-3
r_c^4\right)}{r_+^4 r_c^4 f\left(r_c\right)}\epsilon^{i j k}\partial
_j\beta _k, \label{Current}\eeqs where $j_x(r_+)/r_+^4$
is complicated, and the result can been found in Appendix C of
Ref.~\cite{Bai:2012ci}. Obviously, from the review in the above
section, we need choose new conditions to fix these seven parameters
to find out the bulk viscosity of dual fluid.

Note that, since the boosted transformation relates to the flat metric
or conformal flat metric, the Dirichlet boundary condition
in~(\ref{DirichletB}) can be still imposed, which can keep the boundary induced metric
$\gamma_{\mu\nu}$ conformal to $\eta_{\mu\nu}$ and fix the boundary geometry. Therefore, a way for obtaining the new conditions is
relaxing the conditions in~(\ref{LandauFrame}). In this letter, we
find out that a simple way to obtain the rest equations to fix the seven parameters can
be
\begin{equation}
T^{(1)}_{v v}\neq0,~~T^{(1)}_{v x}=0,~~J_{(1)}^{v}=0,~~
h(r)\equiv0. \label{NewCondition}
\end{equation}
where the underlying consideration is that the gauge choice should be kept, and hence a simple case is relaxing $T^{(1)}_{v v}=0$ in (\ref{LandauFrame}) that relates to the nonzero perturbation of energy density. In fact, from the physical point of view, it is natural and necessary
to turn on the perturbation of the energy
density of fluid, i.e. $T^{(1)}_{v v}\neq0$, since otherwise the compressional effects of the fluid is not
expected to be observed. Note that, after relaxing $T^{(1)}_{v v}=0$ in (\ref{LandauFrame}), one equation will be lacked to fix the seven parameters. We find that this equation can be completed through another simple condition $h(r)\equiv0$, i.e. $h(r_c)=0$ and $h'(r_c)=0$, which can still keep the Dirichlet boundary condition (\ref{DirichletB}). Therefore, after considering the above conditions, the nine
parameters can be finally fixed
\begin{eqnarray}
&&{\Chone}=0,~~{\Chtwo}=0,~~{\Ckone}=0,~~{\Cavone}=0,~~{\Cavtwo}=0,~~C=3\nn\\
&&{\Cktwo}=-\frac{2\partial _i\beta _i
r_c^3}{3\sqrt{f\left(r_c\right)}},~~{\Cthree}=0,~~{\Cfour}=\partial
_v\beta _x r_c^3 \left(-1+\sqrt{f\left(r_c\right)}\right)+\frac{Q
\partial _xQ-\partial _xM r_c^2}{r_c^3 \sqrt{f\left(r_c\right)}}.
\label{NewParameters}
\end{eqnarray}
where $\Chone,$ $\Chtwo$ are from $h(r)\equiv0$. $\Cktwo$, $
\Cavone$ and $ \Cavtwo$ are from requiring $a_v(r_c)=0$, $k(r_c)=0$
and $J_{(1)}^{v}=0$. Two
constants $\CTHREE$, $\CFOUR$ contained in $j_i(r)$ can be fixed by
$T^{(1)}_{v x}=0$ and $j_i(r_c) = 0$.

After inserting (\ref{NewParameters}) into (\ref{BoundaryST})
(\ref{Current}), the non-zero components of stress tensor $T_{\mu\nu}$
are \beqs
    & &T^{(0)}_{v v} = 6\left(1- \sqrt{f\left(r_c\right)} \right) r_c^2, \quad T^{(0)}_{i i} =\frac{-4 M+6\left(1- \sqrt{f\left(r_c\right)}\right) r_c^4}{r_c^2\sqrt{f\left(r_c\right)}},\nn\\
   & &T^{(1)}_{v v}=-2\partial _i\beta _i r_c,~~T^{(1)}_{i j} = \frac{- 2r_+^3\sigma _{ij}}{r_c^2}+\frac{2r_c(2 Q^2-2 M r_c^2-r_c^6)}{3( Q^2-2 M r_c^2+r_c^6)}\partial _i\beta _i,
\eeqs which can be further rewritten in a covariant form
\begin{eqnarray}
T_{\mu \nu}=\rho\,U_\mu U_\nu+p
\Pi_{\mu\nu}-2\eta\sigma_{\mu\nu}-\zeta\theta \Pi_{\mu\nu},
\label{StressTensor1}
\end{eqnarray}
where
\begin{eqnarray}
&&U_{\mu}=r_c u_{\mu},~~
\Pi_{\mu\nu}=\gamma_{\mu\nu}+U_{\mu}U_{\nu}=r_c^2\eta_{\mu\nu}+U_{\mu}U_{\nu},
~~\theta=\nabla_{\mu} U^{\mu}, \nonumber\\
&&\sigma ^{\mu  \nu }\equiv \frac{1}{2} \Pi^{\mu  \alpha } \Pi^{\nu
\beta } \left(\nabla _{\alpha }U_{\beta } +\nabla _{\beta }U_{\alpha
}\right)-\frac{1}{3} \Pi^{\mu  \nu } \nabla _{\alpha }U^{\alpha },
\end{eqnarray}
and the energy density $\rho$, pressure $p$, shear viscosity $\eta$
and bulk viscosity $\zeta$ are
\begin{eqnarray}
&&\rho=2\left(3-3 \sqrt{f\left(r_c\right)} \right) -2\theta,~~~p=\frac{-4 M+2\left(3-
3\sqrt{f\left(r_c\right)}\right)
r_c^4}{r_c^4\sqrt{f\left(r_c\right)}},
~~~\eta=\frac{r_+^3}{r_c^3},\nn\\
&&\zeta=-\frac{2(2 Q^2-2 M r_c^2-r_c^6)}{3( Q^2-2 M
r_c^2+r_c^6)}. \label{ets}
\end{eqnarray}
Note that, the stress tensor $T_{\mu\nu}$
of the dual fluid resides on the spacetime $\gamma_{\mu\nu}=r_c^2\eta_{\mu\nu}$. The bulk viscosity indeed appears in this simple case, and it can be found that the bulk viscosity is always positive between the region $r=r_{+}$ and $r=r_{c}$ by using the Hawking temperature $T_{+}\geq0$ in (\ref{IVTemperature}). In addition, the term
$-2\theta$ in the energy density $\rho$
can be considered as the perturbation of energy density
$\delta\rho$. Since the parameters $C_{j1}, C_{j2}$ related to the
charged current are same, the charged current under the new conditions keeps same as
(\ref{IVfirst order current}).


\section{Conclusion and discussion}

In this paper, we mainly investigate the possibility to find out the
bulk viscosity of dual fluid at the finite cutoff surface via
fluid/gravity correspondence. In Ref~\cite{Bai:2012ci}, it has been
shown that the dual charged fluid can be generalized from infinite boundary to finite cutoff surface.
Although the generalization of dual charged fluid from infinite
boundary to finite cutoff surface via the gravity/fluid
correspondence is straightforward, there are some important
differences between the infinite and finite case. First, the stress
tensor and charged current at the finite cutoff surface can depend
on several undetermined parameters. Excluding the two parameters $\Ckone$ and $C$, one need choose suitable boundary conditions and gauge to fix the other seven parameters, which means
that different dual physics may be corresponded at the finite cutoff
surface. Second, the dual fluid at finite cutoff surface is no
longer conformal, which can be expected from the fact that the conformal symmetry has been broken with a finite
radial coordinate in the bulk. It should be pointed out that although the dual fluid
at finite cutoff surface is non-conformal, but it is usually with
zero bulk viscosity in the fluid/gravity literatures~\cite{Brattan:2011my,Bai:2012ci}. Usually, the nonzero bulk viscosity can also
break the conformal symmetry of dual fluid. Therefore, in this letter we mainly focus on finding out the holographic bulk viscosity at the finite cutoff surface, which is based on~\cite{Bai:2012ci} and uses the gravity/fluid correspondence. Our results show that the nonzero bulk viscosity can indeed be obtained in the dual fluid at the finite cutoff surface after one chooses a new conditions. Moreover, besides the bulk
viscosity term, one can also find a new term related to the perturbation of energy density, which can be expected because the bulk viscosity
is related to the compressional effects of the fluid.

It should be emphasized that in this letter we just consider a simple case to find out the holographic bulk viscosity at the finite cutoff surface, which is achieved by choosing a new simple condition in (\ref{NewCondition}). In fact, we have also investigated another boundary conditions instead of $h(r)\equiv0$ in (\ref{NewCondition}), i.e. relaxing the boundary condition $h'(r_c)=0$. Then, the results can lead to one more physical degree of freedom of the dual fluid , i.e. perturbation of the pressure~\cite{Cai:2011xv}. Concentrating on the holographic bulk viscosity, we just use the simpler case $h(r)\equiv0$ in this letter. However, whether there are other conditions that can be chosen to obtain the nonzero bulk viscosity is still an interesting open question. Moreover, there have been other methods and work to investigate the bulk viscosity~\cite{Buchel:2007mf,Gubser:2008sz,Yarom:2009mw,Eling:2011ms,Buchel:2011yv,Brattan:2011my,Camps:2010br,Emparan:2012be,Emparan:2013ila}. Thus the comparison between our results and their results is necessary for the future work. Note that, if one does not use the conditions in (\ref{LandauFrame}), a general form of $T^{(1)}_{\mu\nu}$ can be found in Ref \cite{Emparan:2013ila}. After comparing with their result (3.52) in \cite{Emparan:2013ila}, we can find that our result (\ref{StressTensor1}) is in fact consistent with their, since the 4-acceleration of $U^a$ in our case is zero, while the perturbation of energy density term in (\ref{StressTensor1}) is just the last term in (3.52) in \cite{Emparan:2013ila}. However, since we discuss different background in the bulk, thus the values of transport coefficients are different. Moreover, as discussed in \cite{Emparan:2013ila}, there is an ambiguity for the extra correction term $g^{(1)}$ in (\ref{IVcorrection}). Our extra correction term $g^{(1)}$ in (\ref{IVcorrection}) is chosen under the gauge in (\ref{gauge}), which is different from their choices in \cite{Emparan:2012be,Emparan:2013ila}. Therefore, it may be another reason to obtain different transport coefficients, which is also needed to be further studied. In addition, our stress tensor of dual fluid $T_{\mu\nu}$ resides on the spacetime $\gamma_{\mu\nu}=r_c^2\eta_{\mu\nu}$, while their results reside just on $\eta_{\mu\nu}$ \cite{Emparan:2012be,Emparan:2013ila}. Since $\gamma_{\mu\nu}$ is conformal to $\eta_{\mu\nu}$, thus the stress tensor of dual fluid resides on $\eta_{\mu\nu}$ usually can be obtained by making a conformal transformation of $T_{\mu\nu}$. However, since the trace of $T_{\mu\nu}$  is usually nonzero, i.e. expressing a non-conformal dual fluid at the finite cutoff surface, thus the stress tensor of dual fluid resides on $\eta_{\mu\nu}$ can not be simply obtained by making a conformal transformation of $T_{\mu\nu}$. Therefore, the conformal symmetry breaking properties of the dual fluid at finite cutoff surface is also an interesting issue to be focused on.


\section{Acknowledgements}

Y.P Hu thanks Profs. Rong-Gen Cai, Bin Chen, Drs. Zhang-Yu Nie, and Jian-Hui Zhang
for useful discussions. Particularly, he thanks a lot for the KITPC's hospitality during his visiting KITPC, Beijing, China. This work is supported by National Natural Science Foundation of China (NSFC) under grant Nos. 11105004, 11175245, 11075206 and
Shanghai Key Laboratory of Particle Physics and Cosmology under
grant No. 11DZ2230700, and partially by grants from NSFC (No.
10821504, No. 10975168 and No. 11035008) and the Ministry of Science
and Technology of China under Grant No. 2010CB833004.


\end{document}